\documentclass[aps,prl,reprint,amsmath,amssymb,floatfix,superscriptaddress]{revtex4-2}
\usepackage[justification=Justified, singlelinecheck = false]{caption}
\usepackage{graphicx}
\usepackage{xcolor}
\usepackage{physics}
\usepackage{hyperref}
\usepackage{upgreek}
\usepackage{nicefrac}

\newcommand{\affA}{Van der Waals-Zeeman Institute, Institute of Physics, University of Amsterdam, 1098 XH Amsterdam, Netherlands}
\newcommand{\affB}{QuSoft, Science Park 123, 1098 XG Amsterdam, the Netherlands}

\newcommand{\affD}{ETH Zürich – PSI Quantum Computing Hub, Paul Scherrer Institut (PSI), 5232 Villigen, Switzerland}
\newcommand{\affE}{Institute for Quantum Electronics, ETH Zürich, Otto-Stern-Weg 1, 8093 Zürich, Switzerland}
\newcommand{\affF}{Quantum Center, ETH Zurich, CH-8093, Switzerland}

\newcommand{\affG}{Department of Physics, University of Trieste, 34127 Trieste, Italy}

\newcommand{\equalcontrib}{\thanks{These authors contributed equally to this work.}}

\begin{document}

\title{Direct observation of the optical Magnus effect with a trapped ion}

\author{Philip Leindecker}
\email{pleindecker@ethz.ch}
\equalcontrib
\affiliation{\affD}\affiliation{\affE}

\author{Louis P.H. Gallagher}
\equalcontrib
\affiliation{\affA}

\author{Edgar Brucke}
\affiliation{\affD}\affiliation{\affE}

\author{Dominique Zehnder}
\affiliation{\affD}\affiliation{\affE}

\author{Luka Milanovic}
\affiliation{\affD}\affiliation{\affE}

\author{Matteo Marinelli}
\affiliation{\affD}\affiliation{\affE}\affiliation{\affG}

\author{Rene Gerritsma}
\affiliation{\affA}\affiliation{\affB}

\author{Robert J.C. Spreeuw}
\affiliation{\affA}\affiliation{\affB}

\author{Jonathan Home}
\affiliation{\affE}\affiliation{\affF}

\author{Cornelius Hempel}
\affiliation{\affD}\affiliation{\affE}\affiliation{\affF}

\date{\today}

\begin{abstract}
    We directly observe and spatially map an optical analog of the Magnus effect, where intrinsic spin-orbit-like coupling of light generates a spin-dependent transverse displacement of the atom–light interaction profile for a $^{40}$Ca$^+$ ion.
    Probed on a quadrupole transition using a tightly focused beam, we observe displacements of the maximum in the profile of the effective interaction by several $100$~nm originating from intrinsic longitudinal electric field components beyond the paraxial approximation. 
    The tight focus of the beam induces additional transverse polarization gradients, which we characterize through a phase-sensitive measurement and spatial maps for different beam configurations.
    The results establish the physical basis of polarization-gradient interactions relevant to optical tweezer-based quantum control.
\end{abstract}

\maketitle

The Magnus effect~\cite{Magnus:1853} describes how a rotating object moving through a medium experiences a force that is orthogonal to both its velocity and its axis of rotation. 
In sports, this effect is widely exploited to send a spinning ball along a curved trajectory, controlled by the direction of rotation. 
An optical analog of the Magnus effect has been predicted for atoms trapped in optical tweezers~\cite{Spreeuw:2020,Spreeuw:2022}, where intrinsic spin-orbit-like coupling of light (Appendix A) gives rise to a spin dependent transverse structure of the atom-light interaction, which is not present in the plane wave approximation. 

A tightly focused Gaussian beam without intrinsic orbital angular momentum (OAM) acquires local OAM and longitudinal electric field components in the nonparaxial description. 
The interplay of the induced OAM and the beam’s spin angular momentum produces spin-orbit-like coupling and an inverted wavefront curvature across the focal plane, generating strong transverse polarization gradients and shifting the maxima of its circular field components off axis~\cite{Zhao:2007,Nieminen:2008,Bliokh:2015,Neugebauer:2018,Eismann:2019}. 
When an atom is placed within the beam, the shift of these field components leads to a transverse displacement of the effective atom-light interaction profile.
The predicted shift is wavelength ($\lambda$) dependent and calculated to lie between $\lambda/2\pi$ and $\lambda/\pi$ for different components of the quadrupole transition~\cite{Gallagher:2026} and $\lambda/2\pi$ for dipole transitions~\cite{Spreeuw:2020,Spreeuw:2022}.
At the beam center, this shift causes the atom to experience an intensity gradient of the light, thereby enabling transverse motional excitations forbidden for a paraxial Gaussian beam, analogous to the interaction of an atom with structured light containing intrinsic OAM~\cite{Schmiegelow:2016,Afanasev:2018,Quinteiro:2017,verde2023trapped,Blodgett:2025,Stopp_ang_mom2022}.
A related polarization dependent spatial dependence of the origin of displacement in light \emph{emission} was previously observed with a trapped ion~\cite{araneda_wavelength-scale_2019}.

Spin-dependent transverse forces can introduce unwanted coupling to the motion of ions positioned at the center of the beam~\cite{Wang:2020,Gallagher:2026,Unnikrishnan:2024,Thompson:2013}, but the same mechanism enables new optical quantum gate schemes based on engineered polarization gradients~\cite{Mazzanti:2023,Cui:2025,Mai:2025}. 
As optical tweezers become increasingly central to atomic physics -- enabling parallel trapping and control of large neutral-atom arrays~\cite{Barredo:2016,Endres:2016,Kaufman:2021,Urech:2022,Bluvstein:2024,Pause:24} and offering new routes to high-fidelity trapped-ion gates~\cite{Olsacher:2020,Mazzanti:2021,Mazzanti:2023,Schwerdt:2024,Schwerdt:2026} -- understanding such intrinsic transverse interactions becomes important. 

In this Letter, we report the first direct observation of the optical Magnus effect in the interaction of a single trapped ion with a tightly focused Gaussian beam. 
Using sub-100-nm spatial resolution in the positioning of laser beams enabled by two crossed acousto-optic deflectors (AODs)~\cite{pogorelov_compact_2021}, we map the spatial profile of coupling strength of all components of the $729$~nm quadrupole transition from $4S_{1/2}\rightarrow\,3D_{5/2}$ in $^{40}$Ca$^+$. 
We resolve the spin-dependent transverse displacement of the interaction profile arising solely from the Gaussian beam's intrinsic longitudinal components~\cite{Spreeuw:2020,Spreeuw:2022}.
We also characterize additional transverse polarization gradients arising from the tight focus of the beam leading to position-dependent selection rules for the quadrupole transition.

\begin{figure*}[t]
    \centering
    \includegraphics[width=\linewidth]{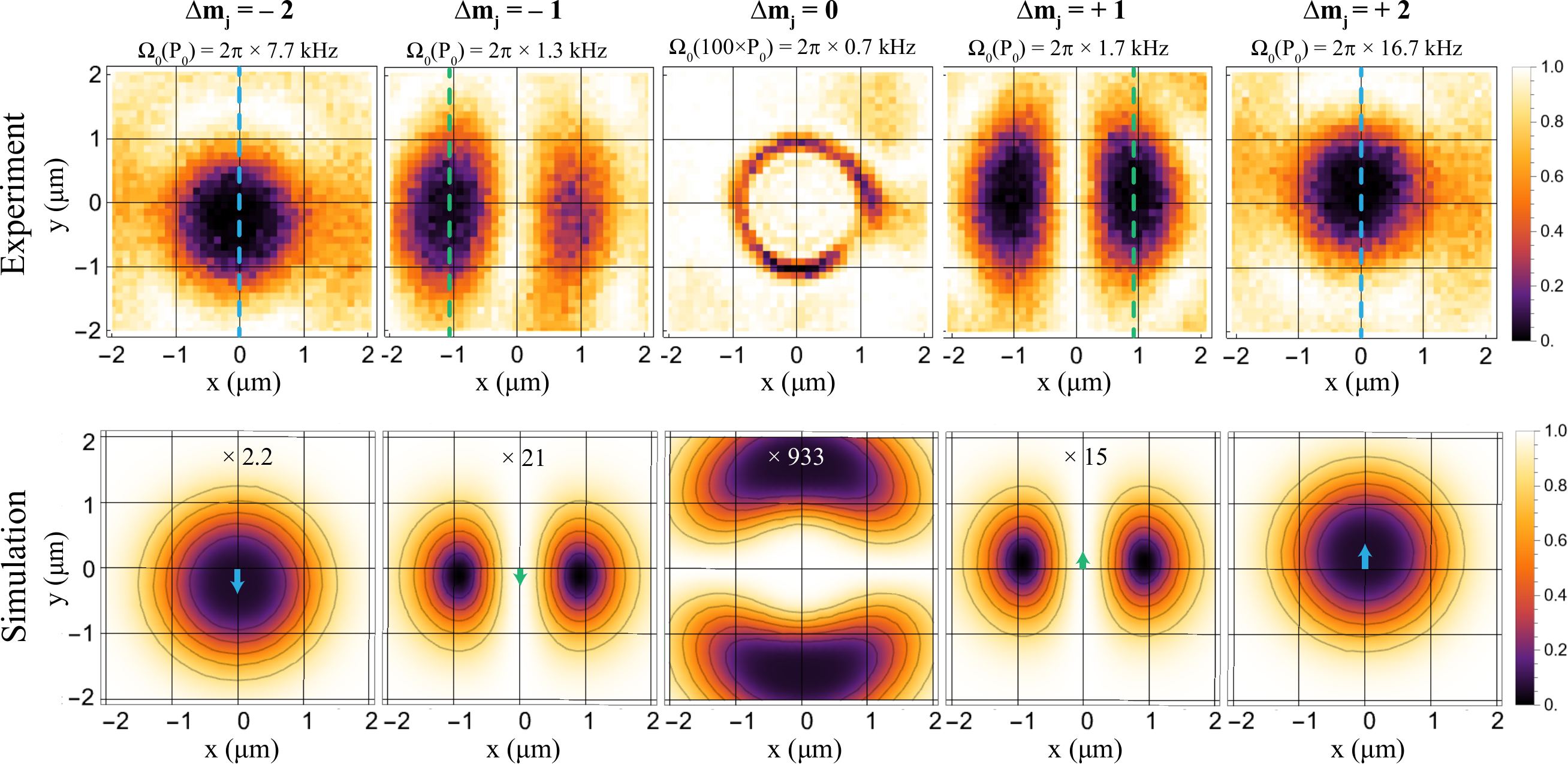}
    \caption{\textbf{Linear polarization.} 
    Spatial profiles of the normalized quadrupole couplings $\ket{4S_{1/2},\,m_j}\rightarrow\, \ket{3D_{5/2},\,m_j+\Delta m_j}$, with $\Delta m_j=0,\pm1\pm2$, for linear polarization along $y$, and $\hat{\boldsymbol{\varepsilon}}\perp\mathbf{B}\perp \boldsymbol{k}$. 
    At each position $(x,y)$, we apply the tweezer for a duration $T_{\text{probe}}$, calibrated to yield a $\pi$ pulse on each transition at the maximum Rabi frequency $\Omega_0$, and then measure the ground state population.
    The experimental maps (top row) are compared to simulated maps (bottom row) normalized to $\Omega_0$ of the $\Delta m_j=+2$ case (Eq.~\eqref{eq:rabi freq}), with scale factors given in each panel.
    Blue and green arrows in the center of the simulated maps highlight the transverse displacement arising from the optical Magnus effect. 
    The $\Delta m_j=\pm1$ and $\Delta m_j=\pm2$ coupling profiles are displaced by approximately $\pm\lambda/2\pi\approx 115$~nm and $\pm\lambda/\pi\approx 230$~nm in the $y$ direction. 
    We resolve the strongly suppressed $\Delta m_j=0$ transition by increasing the probe power $P_0$ by a factor of $\approx 100$. 
    Differences between experiment and simulation likely arise from polarization angle sensitivity and  beam aberrations. 
    Vertical dashed lines mark the cuts for Fig.~\ref{fig:magnus_effect}.}
    \label{fig:linear_pol}
\end{figure*}

The light-matter interaction can be formally described as follows:
the electric field of the tightly focused tweezer beam can be approximated to first order beyond the paraxial limit as~\cite{Novotnoy:2012,Aiello:2015} 
\begin{equation}
    \mathbf{E}(\mathbf{r},t) \approx {\rm Re}\left[\left(\hat{\boldsymbol{\varepsilon}}-\frac{i\left(\varepsilon_{x}x+\varepsilon_{y}y\right)}{z_{0}}\hat{\mathbf{z}}\right)f(\mathbf{r})e^{i\left(kz-\omega t\right)}\right],
\label{eq:E-1st-order}
\end{equation}
where $\hat{\boldsymbol{\varepsilon}} = \varepsilon_x \hat{\mathbf{x}} + \varepsilon_y \hat{\mathbf{y}}$ is the transverse polarization unit vector, $z_0 = \pi w_0^2/\lambda = kw_0^2/2$ is the Rayleigh range and $f(\mathbf{r})$ is the Gaussian beam mode function, defined explicitly alongside the other parameters in Appendix B. 
The Hamiltonian of the quadrupole atom-light interaction is $H_{E2}(\hat{r})=e\hat{r}_i \hat{r}_j \partial_i E_j$ with Einstein's summation implied, $e$ the elementary charge, $\hat{r}_i$ the electron position operator, and \(E_j\) the electric field component from Eq.~\eqref{eq:E-1st-order}, both evaluated at the ion's center-of-mass.
The Hamiltonian components which drive the quadrupole transition $\ket{4S_{1/2},\, m_j}\leftrightarrow\, \ket{3D_{5/2},\, m_j+\Delta m_j}$ with magnetic quantum number $m_j$ and \mbox{$\Delta m_j=0,\pm1,\pm2$} are defined in Appendix B.

In our experiments, we use a single $^{40}$Ca$^+$ ion confined in a linear Paul trap with trap frequencies $(\omega_x,\omega_y,\omega_z)=2\pi\times(1.18,2.38, 2.07)$~MHz, shown schematically in Appendix C~\cite{Brucke:2025}. 
The ion is initialized in the $\ket{4S_{1/2},\,m_j=+1/2}$ Zeeman level and cooled close to the motional ground state by a combination of dark resonance and resolved sideband cooling. 
An optical tweezer beam propagating along the $z$ axis at a wavelength of $729$~nm is tuned to the quadrupole-allowed transitions between the $\ket{4S_{1/2},\,m_j=+1/2}$ state and the different Zeeman levels of the $3D_{5/2}$ state.
The beam is focused to a diameter of $2w_0 \approx 2.6~\upmu$m using a custom NA = 0.4 objective. 
We use two crossed AODs to position the tweezer in the focal plane and then scan it across the ion location to map the spatial dependence of the quadrupole interaction. 
Slow drifts in the ion's position during the measurements are tracked and compensated, as described in Appendix D. 
A magnetic field of $B=0.4$~mT is applied along the $-x$ direction using permanent magnets, resulting in a Zeeman splitting of $\approx 6.72$~MHz within the $3D_{5/2}$ manifold.

For each measurement, the tweezer frequency is adjusted to resonantly drive a selected transition $\ket{4S_{1/2},\,m_j}\leftrightarrow\,\ket{3D_{5/2},\,m_j+\Delta m_j}$. 
After the probe interaction time $T_{\text{probe}}$, we discriminate the final state via state-dependent fluorescence from an additional dipole transition at $397$~nm~\cite{myerson_high-fidelity_2008, burrell_high_2010}.
To extract the transverse interaction profile, each transition is treated independently by normalizing its spatial map to the maximum carrier Rabi frequency $\Omega_0$ measured for that transition.
The ground state population is given by \mbox{$P(x,y)=1-\sin^2\bigl({\Omega_{E2}(x,y) \pi}/{(2\Omega_0)}\bigr)$}, 
where $\Omega_{E2}(x,y)$ is the local quadrupole Rabi frequency. 
This protocol enables direct spatial imaging of the spin-dependent atom–light interaction profile with sub-wavelength resolution.

Fig.~\ref{fig:linear_pol} shows the measured quadrupole coupling strength as a function of tweezer position $(x,y)$ in the focal plane, together with simulated maps calculated using Eq.~\eqref{eq:rabi freq} in Appendix B.
The tweezer polarization $\hat{\boldsymbol{\varepsilon}}$ in this case is vertically oriented along the $y$-axis, enhancing the $\Delta m_j=\pm2$ transitions at the center of the beam through the geometric dependence of the quadrupole interaction~\cite{Roos:2000,Gallagher:2026}. 
To reveal the spatial profiles of the suppressed $\Delta m_j=0$ transitions the optical power was increased by a factor $\approx 100$. The peak Rabi frequencies $\Omega_0$ were extracted from Rabi flops performed at the maximum of each transition, yielding $\Omega_{0}=2\pi\times(7.7,1.3,0.7,1.7,16.7)$~kHz for $\Delta m_j=(-2,\dots,+2)$. 

The coupling profiles in Fig.~\ref{fig:linear_pol} exhibit both spin-dependent transverse displacements due to the optical Magnus effect as well as transverse polarization gradients due to the tight-focusing of the beam. 
The optical Magnus effect causes the $\Delta m_j=\pm1$ and $\Delta m_j=\pm2$ couplings to be shifted in the $y$ direction, as indicated by the green and blue arrows respectively, which we characterize in detail below. 
Note that while the $\Delta m_j=\pm2$ couplings have a Gaussian profile, the $\Delta m_j=\pm1$ couplings are suppressed at the beam center, but the field gradients due to the tight-focusing of the beam result in two distinct lobes. 
As these lobes are due to transverse (rather than longitudinal) electric field gradients they persist even in the paraxial description. 
While the measured coupling profiles show the absolute magnitude of the quadrupole interaction, the two lobes have opposite phases, giving rise to a nonzero field gradient at the beam center, a feature which we explore in more detail later. 
The $\Delta m_j=0$ transition is similarly expected to be driven by transverse field gradients near the edges of the beam. 
However, the experimental data reveal a ring-like coupling pattern. 
This transition is highly sensitive to the angle of polarization; small variations significantly alter the observed pattern. 
An alternative source of imperfection might arise from the ion sitting away from the focal position, however exploring this through numerical simulations we could not recreate the observed differences. 
We therefore think that residual polarization imperfections originating from small misalignments of the waveplate positions or aberrations in the tweezer are the likely cause of the disparity.

\begin{figure}[t!]
    \centering
    \includegraphics[width=\linewidth]{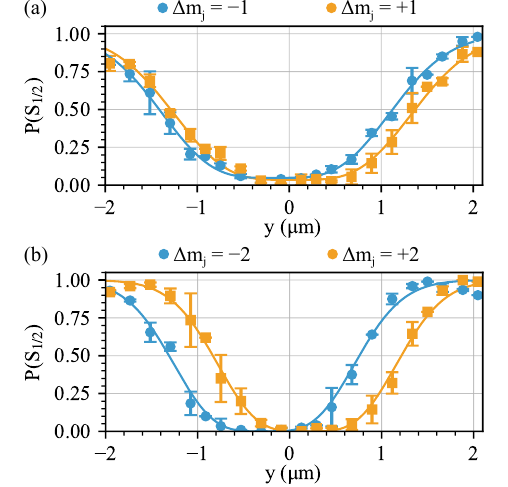}
    \caption{\textbf{Magnus effect displacement.} 
    Relative quadrupole coupling for $\hat{\boldsymbol{\varepsilon}}\perp\mathbf{B}\perp \boldsymbol{k}$ as a function of position $y$ around the maximum coupling indicated by blue and green vertical dashed lines in Fig.~\ref{fig:linear_pol}.
    (a) For $\Delta m_j=\pm1$ we obtain $\Delta y=240(16)$~nm, in agreement with the theoretically predicted value of $\lambda/\pi=232$~nm. 
    (b) For the $\Delta m_j=\pm2$ couplings we find $\Delta y= 463(20)$~nm, consistent with the calculated value $2\lambda/\pi=464$~nm.
    The measured ground state population is binned in steps of $0.2~\upmu$m and error bars indicate the standard deviation within each bin. 
    The solid curves are Gaussian fits through which we find the separation between the minima caused by the optical Magnus effect.}
    \label{fig:magnus_effect}
\end{figure}

\begin{figure*}[t!]
    \centering
    \includegraphics[width=\linewidth]{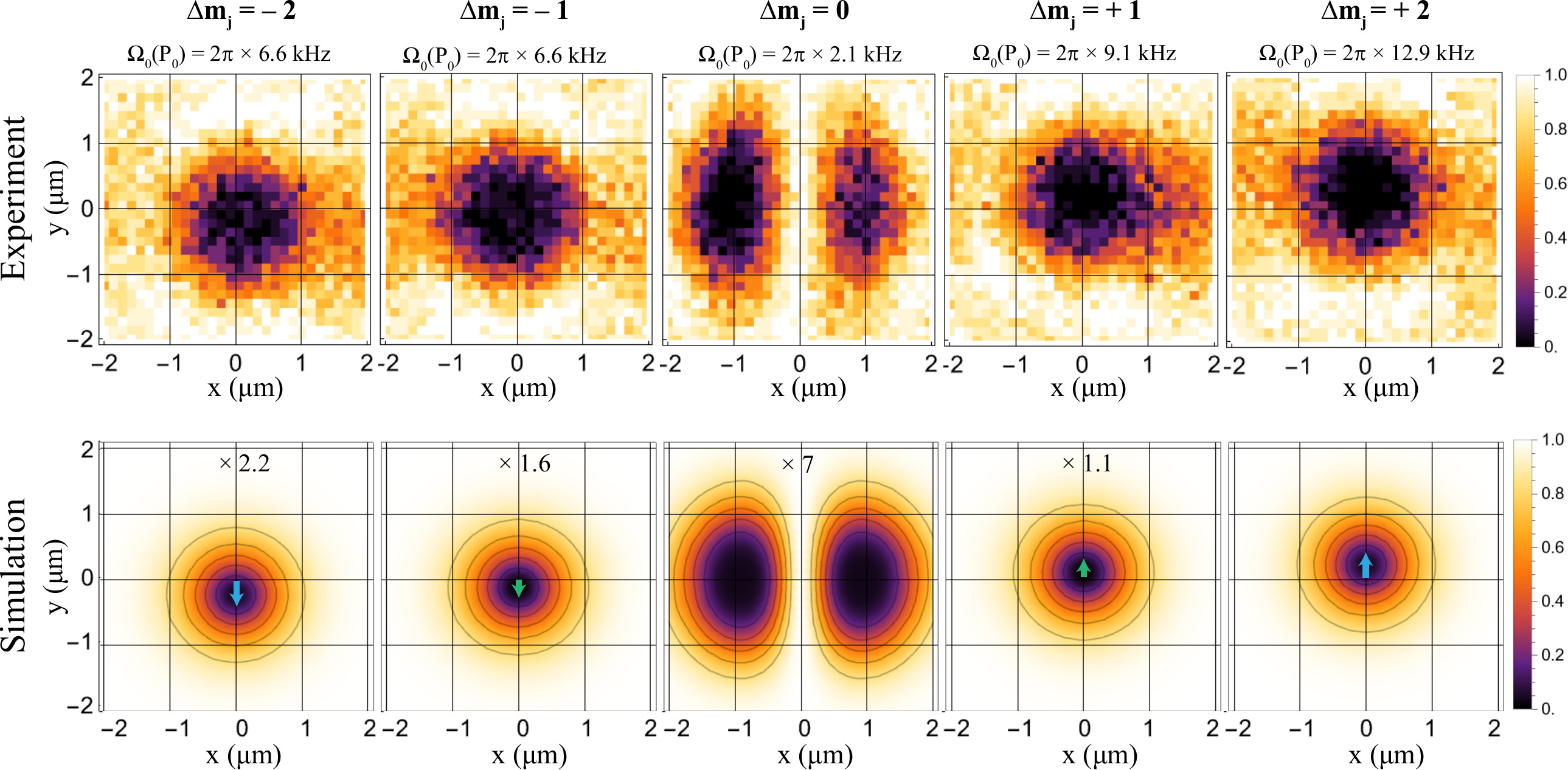}
    \caption{\textbf{Circular polarization.} 
    Spatial profiles of the quadrupole couplings $\ket{4S_{1/2},\,m_j}\rightarrow\, \ket{3D_{5/2},\,m_j+\Delta m_j}$, with $\Delta m_j=0,\pm1\pm2$, for right hand circular polarization. 
    The experimental tweezer maps (top row) are compared to simulated maps (bottom row) calculated using Eq.~\eqref{eq:rabi freq}. 
    The scale factors in each panel give a relative scaling of the maximum Rabi frequency $\Omega_0$ for each transition with respect to the $\Delta m_j=+2$ case. 
    Blue and green arrows highlight the optical Magnus effect displacement. 
    Similar to the linear polarization case, the $\Delta m_j=\pm1$ and $\Delta m_j=\pm2$ couplings are displaced along the $y$ axis by approximately $\pm\lambda/2\pi\approx\pm 115$~nm and $\pm\lambda/\pi\approx\pm 230$~nm, respectively.}
    \label{fig:circ_pol}
\end{figure*}

We now quantify the displacement of the coupling profiles due to the optical Magnus effect in more detail.
Figure~\ref{fig:magnus_effect} shows line cuts along the $y$ axis at the $x$ position which features the maximum coupling for the $\Delta m_j=\pm1$ and $\Delta m_j=\pm2$ transitions. 
The measured Rabi frequencies are fitted using a displaced Gaussian, $\Omega(y)=\Omega_0\exp\left(-\nicefrac{(y-\tilde{y})^2}{w_0^2}\right)$, where $\tilde{y}$ denotes the position of minimum coupling. 
For the $\Delta m_j=\pm1$ transitions, we extract a minimum separation of $\Delta y=240(16)$~nm in the $y$ direction, in agreement with the theoretical prediction of $\lambda/\pi=232$~nm~\cite{Gallagher:2026}. 
Similarly, for $\Delta m_j=\pm2$ we measure a separation of $\Delta y= 463(20)$~nm, which is consistent with the calculated displacement of $2\lambda/\pi=464$~nm. 

The available power enables Rabi frequencies of $\Omega_0\approx 2\pi\times5$~MHz for the $\Delta m_j=+2$ transition. 
For a Gaussian intensity profile centered on the ion at $(x,y)=(0,0)$, this corresponds to transverse gradients of ${\partial\Omega(y)}/{\partial y} =\Omega'\approx 2\pi\times2.7$~kHz/nm, producing a strong spin-dependent transverse force. 
If uncompensated, such gradients can lead to a significant qubit-motion coupling error during single-qubit gates~\cite{Gallagher:2026}. 
However, calibrating the ion's position relative to the coupling profile provides a straightforward means to suppress this effect.

We next set the tweezer to right-hand circular polarization (RHC), $\hat{\boldsymbol{\varepsilon}}=\varepsilon_x\hat{\mathbf{x}}+i\varepsilon_y\hat{\mathbf{y}}$. 
In this configuration, both the $\Delta m_j=\pm1$ and $\Delta m_j=\pm2$ transitions are enhanced by the geometric dependence of the quadrupole coupling. 
The measured spatial profiles of the couplings are shown in Fig.~\ref{fig:circ_pol}. 
As in the linearly polarized case, we observe a displacement along the $y$ axis for both $\Delta m_j=\pm1$ and $\Delta m_j=\pm2$ couplings, consistent with the optical Magnus effect. 
From Gaussian fits to the data we obtain a separation of $151(21)$~nm for $\Delta m_j=\pm1$ transitions, compared to the predicted $\lambda/\pi=232$~nm. 
The $\Delta m_j=\pm2$ minima are separated by $505(24)$~nm, compared to the predicted $2\lambda/\pi=464$~nm. 
The somewhat larger deviation from simulation, compared to the linear polarization case, likely arises from a combination of reduced spatial resolution and sampling, as well as imperfect polarization resulting from stronger residual mechanical and thermal drifts in the addressing and imaging system after in-situ calibration during that particular measurement.
Finally, the $\Delta m_j=0$ transition exhibits two lobes due to the field gradients on the sides of the beam. 

Changing from right- to left-hand circular polarization does not change the coupling profiles as it merely introduces a global phase shift between the $x$ and $y$ components of the beam and does not alter the gradients. 
By contrast, in simulation we see that inverting the magnetic field direction along $x$ reverses the direction of the displacements along $y$.

\begin{figure}[t]
    \centering
    \includegraphics[width=\linewidth]{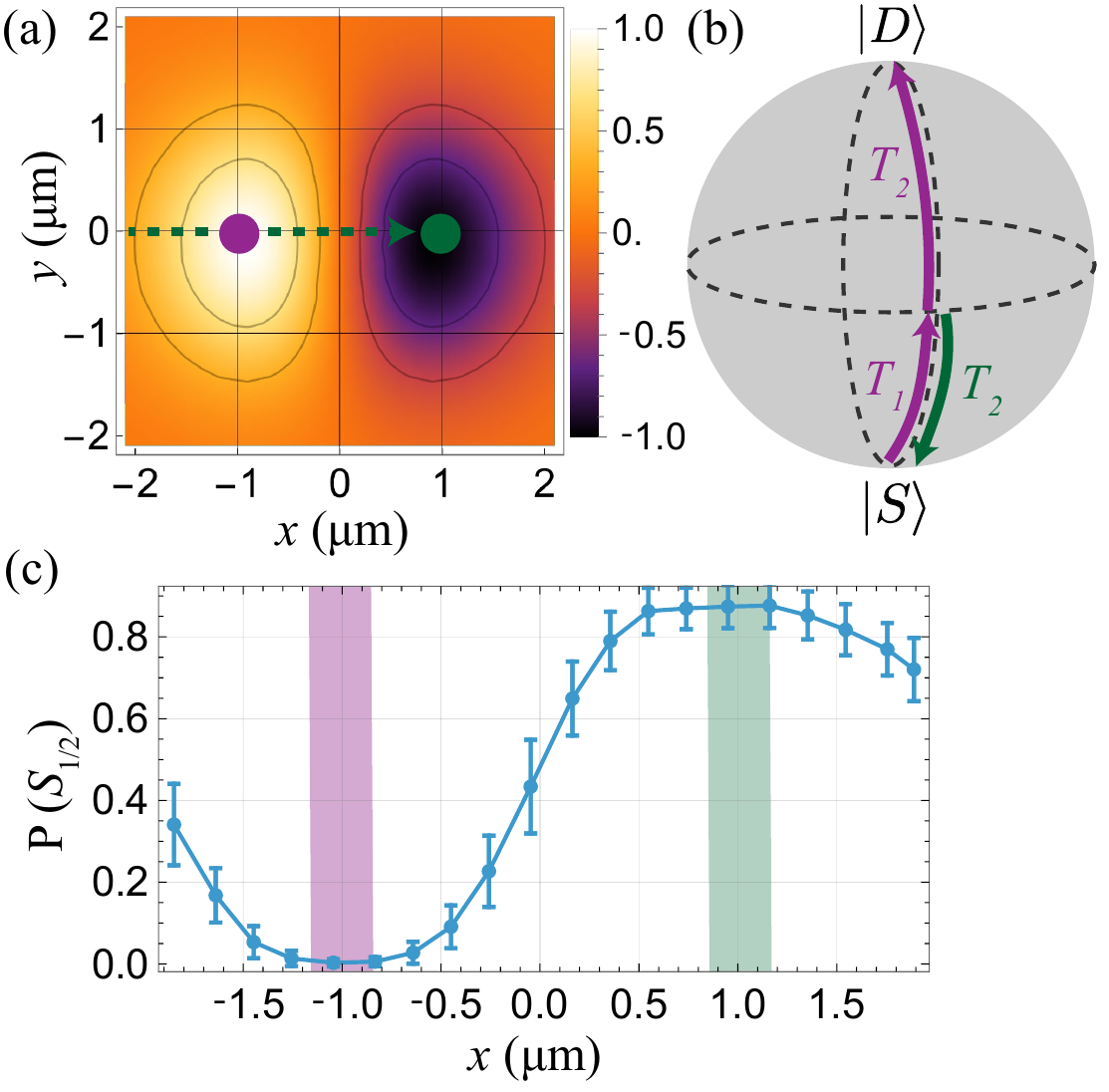}
    \caption{\textbf{Phase of the quadrupole coupling.} 
    (a) Signed Rabi frequency of the $\Delta m_j=+1$ transition with $\hat{\boldsymbol{\varepsilon}}\perp\mathbf{B}\perp \boldsymbol{k}$ as a function of tweezer position $x$, showing that the two lobes have opposite phase. 
    (b) Pulse-sequence schematic on a Bloch sphere for detecting this phase: an initial $\pi/2$ pulse ($T_1$) is applied with the left lobe centered on the ion, followed by a second $\pi/2$ pulse ($T_2$) of equal duration at a tweezer position shifted along $x$. 
    If the second pulse is carried out on the left lobe, it adds to an overall $\pi$ pulse; on the right lobe, it cancels the first pulse. 
    (c) Measured ground state population versus tweezer position along $x$. 
    The data are binned in $0.2~\upmu$m steps and error bars show the standard deviation within each bin. 
    The shaded regions mark the positions of the two lobes.}
    \label{fig:rabi_freq_sign}
\end{figure}

The double lobe structure observed for some configurations in Figs.~\ref{fig:linear_pol} and \ref{fig:circ_pol} arises from transverse field gradients of opposite sign, which we illustrate in Fig.~\ref{fig:rabi_freq_sign} (a) by showing the signed Rabi frequency (Eq.~\eqref{eq:rabi freq}), rather than its magnitude, as a function of $x$ and $y$. 
This implies that at the center of the beam there is a field gradient. 
Intuitively, the curvature of the beam on either side of the tweezer has an opposite sign (corresponding to a $\boldsymbol{k}$ vector rotation), and thus the electric field gradient driving the quadrupole transition does as well. 

To experimentally confirm this phase difference, we implement the $\pi/2$ pulse sequence illustrated in Fig.~\ref{fig:rabi_freq_sign}(b). 
We first position the tweezer so that the left lobe of the $\Delta m_j = +1$ transition illustrated in Fig.~\ref{fig:linear_pol} is centered on the ion.
Then we apply a pulse ($T_1$) of duration $t=\pi/(2\Omega_0)$, where $\Omega_0$ is the maximum Rabi frequency in the center of the left lobe. 
We then shift the tweezer horizontally along the $x$ axis using the AODs and apply a second pulse ($T_2$) of identical duration. 
When this second pulse again addresses the left lobe, the operations add to a full $\pi$ pulse, transferring all $4S_{1/2}$ ground state population to the $3D_{5/2}$ excited state. 
In contrast, when the second pulse addresses the right lobe, its opposite phase cancels the first $\pi/2$ pulse. 
The resulting ground state population as a function of tweezer position $x$ shown in Fig.~\ref{fig:rabi_freq_sign}(c), directly reveals the sign change of the Rabi frequency and the corresponding gradient at position $x=0$.

This field gradient can be exploited to generate a transverse force without coupling to the carrier transition. 
For example, simulating the $\Delta m_j=+1$ transition for $P_0=10~\upmu\textrm{W}$ and $2w_0=2.6~\upmu$m yields a vanishing carrier coupling strength $\Omega_0\rightarrow0$ and a gradient of $\Omega'=2\pi\times76$~Hz/nm at the beam center. 
This gradient magnitude is comparable to that used in Ref.~\cite{Cui:2025} to implement a two-qubit gate in $t\approx 400~\upmu$s and can be increased by raising the optical power. 
Simulations further show that a $1^{\circ}$ deviation from the ideal $\varepsilon\perp\mathbf{B}$ configuration shifts the zero point of the $\Delta m_j=\pm1$ coupling by $\Delta x\approx \pm 130$~nm. 
Consequently, polarization deviations, optical aberrations, and beam-pointing drifts may present challenges for maintaining alignment and stability in experiments.

We have observed the optical Magnus effect on the 729~nm quadrupole transition in $^{40}$Ca$^+$, manifest as a transverse displacement of the atom-light coupling with respect to the beam propagation. 
By characterizing the effect for different polarization configurations, we find that the displacement approaches $\pm\lambda/2\pi$ for $\Delta m_j=\pm1$ and $\pm\lambda/\pi$ for $\Delta m_j=\pm2$, in the direction orthogonal to $\mathbf{B}$, and the resulting field gradients at the beam center may introduce errors if unaccounted for. 
While the optical Magnus effect arises due to the longitudinal field components, the tight focusing of the transverse field induces polarization gradients which we also observe. 
In particular, the $\Delta m_j=\pm1$ transitions exhibit vanishing carrier coupling and a non-zero field gradient at the beam center, making them especially promising for implementing optical two-qubit gates~\cite{Cui:2025,Mai:2025}. 
A key advantage is that the induced forces are purely transverse -- perpendicular to the propagation direction, thus simplifying their use compared with typical M{\o}lmer-S{\o}renson schemes, which rely on the axial projection of the beam's $\boldsymbol{k}$ vector. 

\textit{Acknowledgments}.---We thank Matteo Mazzanti, Zeger Ackerman and Alfredo Ricci-V\'{a}squez for helpful feedback on the manuscript. 
This work was supported by the ETH Zurich–PSI Quantum Computing Hub; the Intelligence Advanced Research Projects Activity (IARPA) and the Army Research Office under the Entangled Logical Qubits program through Cooperative Agreement No.~W911NF-23-2-0216; the Netherlands Organization for Scientific Research (Grant Nos.~680.91.120, VI.C.202.051, and 680.92.18.05); the Dutch Research Council (Grant No.~OCENW.M.22.403); and the Horizon Europe programme HORIZON-CL4-2021-DIGITAL-EMERGING-01-30 via project 101070144 (EuRyQa).

\bibliography{paper_bib}

\appendix

\section{\large{End Matter}}

\section{Appendix A: Spin-orbit-like coupling of light}
\label{sec:spin-orbit-like-coupling}

The term spin-orbit-like coupling of light~\cite{vuong_electromagnetic_2010, zhao_spin--orbital_2007, marrucci_optical_2006} is used here to denote the intrinsic interdependence between the polarization $\hat{\epsilon}$ and the spatial structure of a classical electromagnetic field describing the light of the tweezer (see Eq.~\ref{eq:E-1st-order}). 
In this context, right- and left-handed circular polarizations are described as spin-like degrees of freedom of the light field, whereas the spatially varying phase and amplitude of the field give rise to orbital-like angular momentum (OAM) associated with the transverse field structure. 
Beyond the paraxial approximation, these two properties are no longer separable, leading to a polarization-dependent spatial field distribution. 
This coupling arises entirely within a classical field description and should be clearly distinguished from coupling between true spin and orbital angular momentum of atomic degrees of freedom.

\section{Appendix B: Light-matter interaction beyond the paraxial approximation}
\label{sec:light-matter-interaction}

The electric field of the tightly focused tweezer beam near the focal plane is approximated to first order beyond the paraxial limit as~\cite{Novotnoy:2012,Aiello:2015} 
\begin{equation}
    \mathbf{E}(\mathbf{r},t) \approx {\rm Re}\left[\left(\hat{\boldsymbol{\varepsilon}}-\frac{i\left(\varepsilon_{x}x+\varepsilon_{y}y\right)}{z_{0}}\hat{\mathbf{z}}\right)f(\mathbf{r})e^{i\left(kz-\omega t\right)}\right],\nonumber
\end{equation}
where $\hat{\boldsymbol{\varepsilon}} = \varepsilon_x \hat{\mathbf{x}} + \varepsilon_y \hat{\mathbf{y}}$ is the transverse polarization unit vector, $z_0 = \pi w_0^2/\lambda = kw_0^2/2$ is the Rayleigh range, $k$ is the wavenumber and $\omega$ the angular frequency. 
The Gaussian beam mode function is 
\begin{align*}
f(\mathbf{r}) = E_0\frac{w_{0}}{w(z)}\exp\left(-\frac{\rho^{2}}{w^{2}(z)}+ik\frac{\rho^{2}}{2R(z)}+i\psi(z)\right).
\end{align*}
where $\rho=\sqrt{x^2+y^2}$, $w(z)=w_0\sqrt{1+(z/z_0)^2}$ is the beam waist, $R(z)=(z^2+z_0^2)/z$ is the wavefront radius of curvature, and $\psi(z)=-\arctan(z/z_0)$ is the Gouy phase.

The electric field gradients that drive the quadrupole transition $\ket{4S_{1/2},\, m_j}\rightarrow\, \ket{3D_{5/2},\, m_j+\Delta m_j}$ with magnetic quantum number $m_j$ and \mbox{$\Delta m_j=0,\pm1,\pm2$} are~\cite{Weissbluth:2012,Gallagher:2026} 
\begin{align*}
    (\nabla\mathbf{E})^{(2)}_{\pm2}&=\frac{1}{2}(\partial_x\pm i\partial_y)(E_x\pm iE_y),\\
    (\nabla\mathbf{E})^{(2)}_{\pm1}&=\frac{1}{2}\left[\mp \partial_z(E_x\pm iE_y)\mp (\partial_x\pm i\partial_y)E_z\right], \textrm{ and}\\
    (\nabla\mathbf{E})^{(2)}_{0}&=\frac{\sqrt{6}}{2}\partial_z E_z. 
\end{align*}
The quadrupole Rabi frequency between the ion's ground state $\ket{J_g,m_g}$ and excited state $\ket{J_e,m_e}$ is
\begin{equation}
\Omega_{E2,m_g,m_e}=\frac{e a_0^2}{\hbar}Q_{\rm red}C_{J_g,m_g,2,q}^{J_e,m_e}(-1)^q(\tilde{\nabla}\mathbf{E})^{(2)}_{-q},
\label{eq:rabi freq}
\end{equation}
where $Q_{\text{red}}=\frac{\mel{J_e}{|Q^{(2)}|}{J_g}}{\sqrt{2J_e+1}}$ is the reduced quadrupole moment in atomic units~\cite{Kreuter:2005}, $C_{J_g,m_g,2,q}^{J_e,m_e}$ is the Clebsch-Gordon coefficient and $a_0$ is the Bohr radius. 
Here, $\Tilde{\nabla}$ denotes the field gradients rotated to the quantization frame set by the magnetic field $\mathbf{B}$. The resulting spatial dependence of the Rabi frequency calculated from Eq.~\eqref{eq:rabi freq} is compared to experimental measurements in Fig.~\ref{fig:linear_pol}.

\section{Appendix C: Experimental apparatus}
\label{sec:experimental-apparatus}

\begin{figure}[h!]
    \centering
    \includegraphics[width=0.9\linewidth]{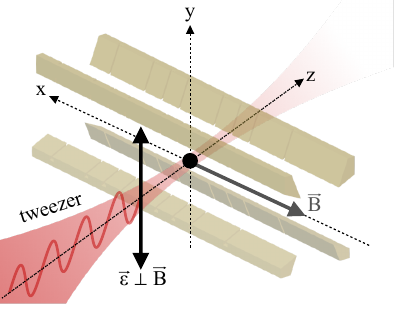}
    \caption{\textbf{Experiment layout.} 
    A laser beam is tightly focused on a trapped $^{40}$Ca$^+$ ion confined in a linear Paul trap. 
    The light of this optical tweezer at $\lambda = 729$~nm is tuned to the narrow $4S_{1/2}\leftrightarrow\,3D_{5/2}$ quadrupole transition. 
    The polarization vector $\hat{\boldsymbol{\varepsilon}}$ illustrated here (with $\hat{\boldsymbol{\varepsilon}} \perp \mathbf{B}$) is the configuration used for the measurements in Fig.~\ref{fig:linear_pol}.}
    \label{fig:experiment_layout}
\end{figure}

Our experimental apparatus operates at room temperature in UHV featuring a monolithic segmented 3D linear Paul trap~\cite{Brucke:2025}. 
A schematic of the experimental setup is shown in Fig.~\ref{fig:experiment_layout}. 
The tweezer propagates along the $z$ axis through the linear Paul trap, and the tweezer scans across positions ($x,y$) to map out the quadrupole interaction profile shown in Figs.~\ref{fig:linear_pol} and \ref{fig:circ_pol} using two crossed acousto-optic deflectors (AODs). 
While the AODs move the tweezer, a double-pass acousto-optical modulator (AOM) setup (positioned before the AODs) keeps the laser frequency on resonance with the transition. 
The magnetic field is oriented along the $-x$ axis. 
A two-layer Mu-metal magnetic shield encloses both the vacuum chamber that houses the ion trap and the surrounding beam delivery optics, including the objective and AOD path, helping to reduce magnetic field noise as well as potential air fluctuations in the beam paths.

The polarization of the tweezer beam can be configured arbitrarily by a combination of a quarter- and a half-waveplate located in the beam path after the AODs. 
In order to optimize the polarization to the two relevant cases (linear and circular) we use a polarimeter in-situ before the objective. 
For the linear case we further fine-tune the polarization with measurements on the ion.

\section{Appendix D: Ion drift correction}
\label{sec:ion-drift}

\begin{figure}[h!]
    \centering
    \includegraphics[width=0.9\linewidth]{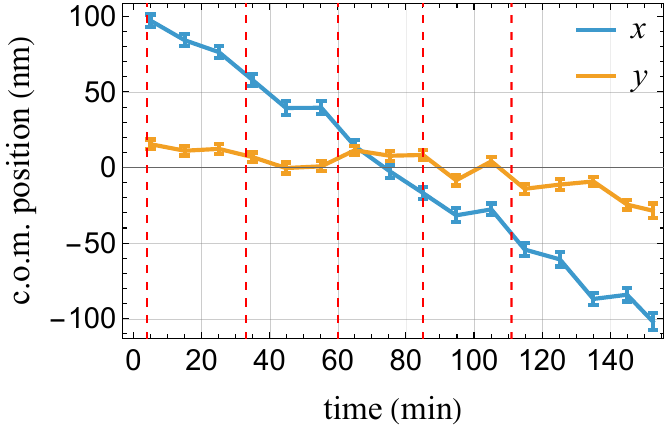}
    \caption{\textbf{Ion drifts.} 
    Drift of the center-of-mass (c.o.m.) position of the ion along $x$ and $y$, measured by a separate EMCCD. 
    The red dashed lines indicate measurement times for each of the couplings plotted in Fig.~\ref{fig:linear_pol}, which each took approximately 25 minutes. 
    Data were binned in time steps of $10$~minutes and the error bar gives the standard error from the mean.}
    \label{fig:drifts}
\end{figure}

To control for drifts in the ion position during experiments, the ion's position was recorded every $100$~ms by a separate EMCCD, and averaged over 10 minutes to give the center-of-mass position. 
We account for drifts in our analysis by recalibrating the position of each dataset with the averaged center-of-mass position at the start of each measurement. 
The drifts during the measurement run for Fig.~\ref{fig:linear_pol} are shown in Fig.~\ref{fig:drifts}, with the start time of each measurement marked with red-dashed lines. 
Drifts during each measurement are of the order $10$~nm along $y$ (the direction in which we measure the optical Magnus effect), which approaches the limits set by the zero-point motion of the ion.

\clearpage

\end{document}